\documentclass[
twocolumn,
 showpacs,amsmath,amssymb]{revtex4}
\usepackage{graphicx}
\usepackage{epstopdf}
\usepackage{bm}
\usepackage{subfigure}
\usepackage[latin1]{inputenc}
\usepackage{rotating}
\usepackage{longtable}
\usepackage{float}
\usepackage[normalem]{ulem}
\usepackage{dcolumn}% Align table columns on decimal point
\usepackage{bm}% bold math
\usepackage{color}

\usepackage{braket}

\begin{document}

\preprint{APS/123-QED}

\title{Doppler-free spectroscopy on Cs D$_1$ line with a dual-frequency laser}

\author{Moustafa Abdel Hafiz$^1$, Gregoire Coget$^1$, Emeric de Clercq$^2$ and Rodolphe Boudot$^1$}

\affiliation{$^1$FEMTO-ST, CNRS, UFC, 26 chemin de l'\'epitaphe 25030 Besan\c{c}on cedex, France \\ LNE-SYRTE, Observatoire de Paris, PSL Research University, CNRS, Sorbonne Universit\'es, UPMC Univ. Paris 06, 61 avenue de l'Observatoire, 75014 Paris, France.}

\date{\today}

%\textcolor{blue}{}

\begin{abstract}
We report on Doppler-free laser spectroscopy in a Cs vapor cell using a dual-frequency laser system tuned on the Cs D$_1$ line. Using counter-propagating beams with crossed linear polarizations, an original sign-reversal of the usual saturated absorption dip and large increase in Doppler-free atomic absorption is observed. This phenomenon is explained by coherent population trapping (CPT) effects. The impact of laser intensity and light polarization on absorption profiles is reported in both single-frequency and dual-frequency regimes. In the latter, frequency stabilization of two diode lasers was performed, yielding a beat-note fractional frequency stability at the level of $3 \times 10^{-12}$ at 1 s averaging time. These performances are about an order of magnitude better than those obtained using a conventional single-frequency saturated absorption scheme.
\end{abstract}

%\pacs{}

\maketitle

\begin{figure}[h!]
\centering
\includegraphics[width=0.95\linewidth]{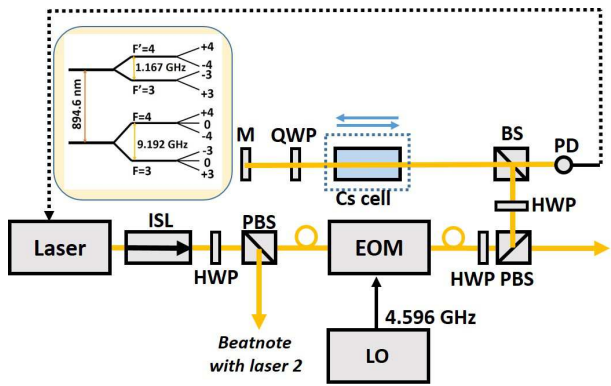}
\caption{Experimental setup for Doppler-free Cs D$_1$ line spectroscopy. EOM: electro-optic modulator, ISL: optical isolator, HWP: half-wave plate, PBS: polarizing beam splitter, BS: beam splitter, QWP: quarte-wave plate, M: mirror, PD: photodiode, LO: local oscillator. The Cs D$_1$ line energy structure is reminded on the top left. }
\label{fig:setup}
\end{figure}

In gas cell experiments, saturated absorption spectroscopy or Doppler-free spectroscopy \cite{Schawlow, Letokhov:1976} is an elegant technique to circumvent Doppler broadening and to allow the detection of natural-linewidth resonance dips in the bottom of absorption profiles. This method, frequently used to stabilize the frequency of a laser to a particular atomic line, is applied in atomic frequency standards \cite{VanierAudoin}, magnetometers \cite{Romalis:Magneto}, laser-cooling experiments \cite{Chu:RMP:1998} or atom interferometry. Laser fractional frequency stability in the 10$^{-13}$$-$10$^{-11}$ range at 1 s integration time can be obtained by combining saturated absorption techniques and narrow-linewidth diode lasers \cite{Rovera:RSI:1994, Affolderbach:RSI:2005, Baillard:OC:2006, Liu:IM:2012, Liang:AO:2015}. In a common saturated absorption scheme, two counter-propagating laser beams of same frequency, a pump-beam and a probe beam, derived from a single laser beam, overlap one another in an atomic vapor cell. Since both beams have the same frequency, the Doppler effect brings different velocity groups into resonance with each beam. On resonance, only atoms presenting zero velocity along the light propagation axis actually experience the two beams with the same laser frequency. If the pump-beam intensity is high enough, the ground state is depleted and therefore the absorption of the probe beam is reduced compared to the case without pump beam. In this case, a so-called Lamb dip, with lorentzian profile, appears at the resonance frequency of the atomic transition in the bottom of a gaussian Doppler-broadened absorption profile. The full-width at half maximum (FWHM) of the Doppler-free Lamb dip can be at low laser intensity as small as the natural linewidth of the atomic transition (4.6 MHz for the Cs D$_1$ line). This process is relatively easy to understand in the case of a simple two-level atom with a single-frequency laser system. In real experience, alkali atoms exhibit a complex multi-level energy structure with hyperfine structure splitting and numerous Zeeman sub-levels, especially in the Cs case (see Fig. \ref{fig:setup}). Such multi-level structures can lead, under appropriate conditions, to a sign reversal of saturation resonances \cite{Pappas:1980, Wynands:APB:1994, McFerran:Arxiv:2016}.\\
In the present article, we report on a Doppler-free laser frequency stabilization setup on Cs D$_1$ line using a coherent bi-chromatic optical field. This setup was initiated in the frame of the development of a high-performance Cs vapor cell atomic clock based on coherent population trapping (CPT) \cite{MAH:JAP:2015}. In many Cs CPT clocks, including chip-scale atomic clocks \cite{Knappe:2007}, the laser is modulated at 4.6 GHz (half of the Cs atom clock frequency). In this configuration, both first-order optical sidebands induce the CPT process and the laser carrier frequency needs to be frequency-stabilized 4.6 GHz away from Cs resonance transition. In CSACs, the laser frequency is currently stabilized onto the bottom of a homogeneously broadened optical line detected in the buffer-gas filled CPT cell, leading to modest laser frequency stabilization. In a high-performance CPT clock \cite{MAH:JAP:2015}, enhanced stabilization of the laser frequency is required. In that sense, we investigated the possibility to perform Doppler-free spectroscopy in a Cs vapor cell with a dual-frequency laser field. Here, in this two-frequency configuration, two different transitions are involved in the saturated absorption scheme, $|F = 3\rangle \rightarrow |F' = 4\rangle$ and $|F = 4\rangle \rightarrow |F' = 4\rangle$, where $F(F')$ are the hyperfine quantum numbers of the ground (excited) states, respectively. Using counter-propagating beams with crossed linear polarizations, the detection of high-contrast natural-linewidth reversed dips is demonstrated. The impact of polarization and laser intensity on absorption profiles is shown, supported by a qualitative theoretical approach. We demonstrate that the dual-frequency configuration allows to enhance greatly the dip-based frequency discriminator slope and the laser frequency stabilization compared to the usual single-frequency Doppler-free spectroscopy.\\
Figure \ref{fig:setup} shows the experimental setup. A single laser beam is and reflected back through a Cs vapor cell to produce both pump and probe beams.

\begin{figure}[h!]
\centering
\includegraphics[width=0.95\linewidth]{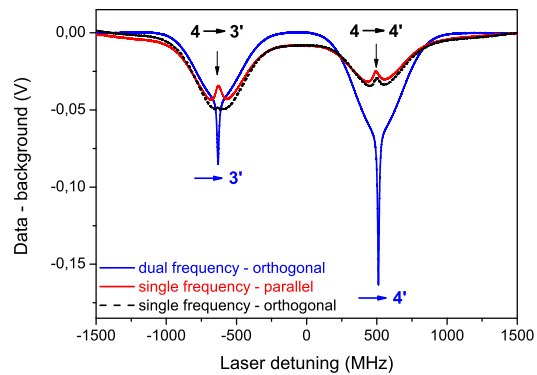}
\caption{Spectroscopy signals detected in the Cs vapor cell in single-frequency (parallel or orthogonally polarized counterpropagating beams) and dual-frequency (crossed polarizations) regimes. For clarification of the figure, a background value of 0.45 V was subtracted to each curve. 3, 4, 3' and 4' stand for $|F  =3 \rangle$, $|F = 4\rangle$, $|F' =3 \rangle$ and $|F' = 4\rangle$ respectively.}
\label{fig:spectres}
\end{figure}

The laser source is a 1-MHz Distributed Feedback (DFB) diode laser tuned on the Cs D$_1$ line at 894.6 nm. An optical isolator is used to prevent optical feedback. A pigtailed Mach-Zehnder intensity electro-optic modulator (MZ EOM), driven at 4.596315 GHz by a low noise microwave frequency synthesizer, allows the generation of two first-order optical sidebands frequency-split by 9.192631 GHz. The optical carrier suppression can be actively stabilized thanks to the technique presented in \cite{Liu:PRA:2013}. The contribution of second-order and third-order optical sidebands, rejected at a level 25 to 35 dB lower, is neglected in the article. A fiber collimator is used to extract a free-space collimated laser beam with a diameter of 1.5 mm from the EOM output. A fraction of the laser power is used to perform the Doppler-free spectroscopy setup. The laser beam is sent and reflected back in a 2-cm-diameter and 2-cm long Cs vapor cell. A quarter wave plate can be adjusted such that propagating and counter-propagating beams are linearly and parallel or mutually orthogonally polarized. The power of the reflected beam is measured by a photodiode. All tests were performed at ambient temperature (about 22.5 $^{\circ}$C). The cell is surrounded by a mu-metal magnetic shield. No static magnetic field is applied. \\
Figure \ref{fig:spectres} shows typical transmission signals detected by the photodiode (PD in Fig. \ref{fig:setup}) in the single-frequency configuration and the dual-frequency regime. The total laser power incident in the cell is 400 $\mu$W in all cases. In the single-frequency case, the transition doublet $|F = 4\rangle \rightarrow |F' = 4\rangle$ and $|F = 4\rangle \rightarrow |F' = 3\rangle$ is shown. With parallel polarizations, the usual narrow Lamb dip, with lorentzian profile, appears in the bottom of both Doppler-broadened absorption gaussian-profile lines. With crossed polarizations, the dip is totally vanished in the bottom of the $|F = 4\rangle \rightarrow |F' = 3\rangle$ transition. In the dual-frequency regime, with crossed polarizations, a significant sign-reversal of the dip and large increase in Doppler-free atomic absorption is observed, especially at high laser power.\\
The dip reversal can be explained by CPT effects which are cancelled for counter-propagating waves. We give here a simple qualitative explanation. A detailed theoretical investigation should take into account all the 25 involved Zeeman sub-levels and can be only numerically computed.

When a multilevel atom is irradiated by a wave connecting in a coherent way two sub-levels of the ground state to a same excited level, forming a $\Lambda$ scheme, the atom can be placed in a light-uncoupled linear superposition of both sub-levels. The atom no longer absorbs the light in this so-called dark state \cite{Arimondo:1996}. A linearly polarized monochromatic wave propagating along the quantification $z$-axis is $\sigma$ polarized. In null magnetic field and at optical resonance, a Zeeman sub-level of the excited state $\ket{e,m}$ will then be linked to two sub-levels of the ground state $\ket{g,m-1}$ and $\ket{g,m+1}$. These conditions allow the existence of a dark state that will disappear for increasing magnetic field, as soon as the Zeeman splitting is larger than the CPT resonance linewidth. This effect is known as the non-linear Hanle effect \cite{Beverini:1985}. An example of full density matrix numerical calculation is reported in \cite{Clercq:1989}. According to the linear polarization along the $x$-axis or $y$-axis, the dark state is given by:
\begin{eqnarray}
\begin{split}
&\ket{ds_{x}}&=&d_{e,m;g,m+1}\ket{g,m-1}+d_{e,m;g,m-1}\ket{g,m+1},\\
&\ket{ds_{y}}&=&d_{e,m;g,m+1}\ket{g,m-1}-d_{e,m;g,m-1}\ket{g,m+1},
\end{split}
\label{eq:dsxy}
\end{eqnarray}
where $d_{e,m;g,m\pm1}$ are the dipole moments of the involved transitions. Equal field amplitudes are assumed. Note that both dark states are orthogonal. This means that the atoms driven in a dark state by a polarized wave will absorb the perpendicularly polarized wave (bright state).
\begin{figure*}[t!]
\centering
\subfigure[]{\includegraphics[width=0.3\linewidth]{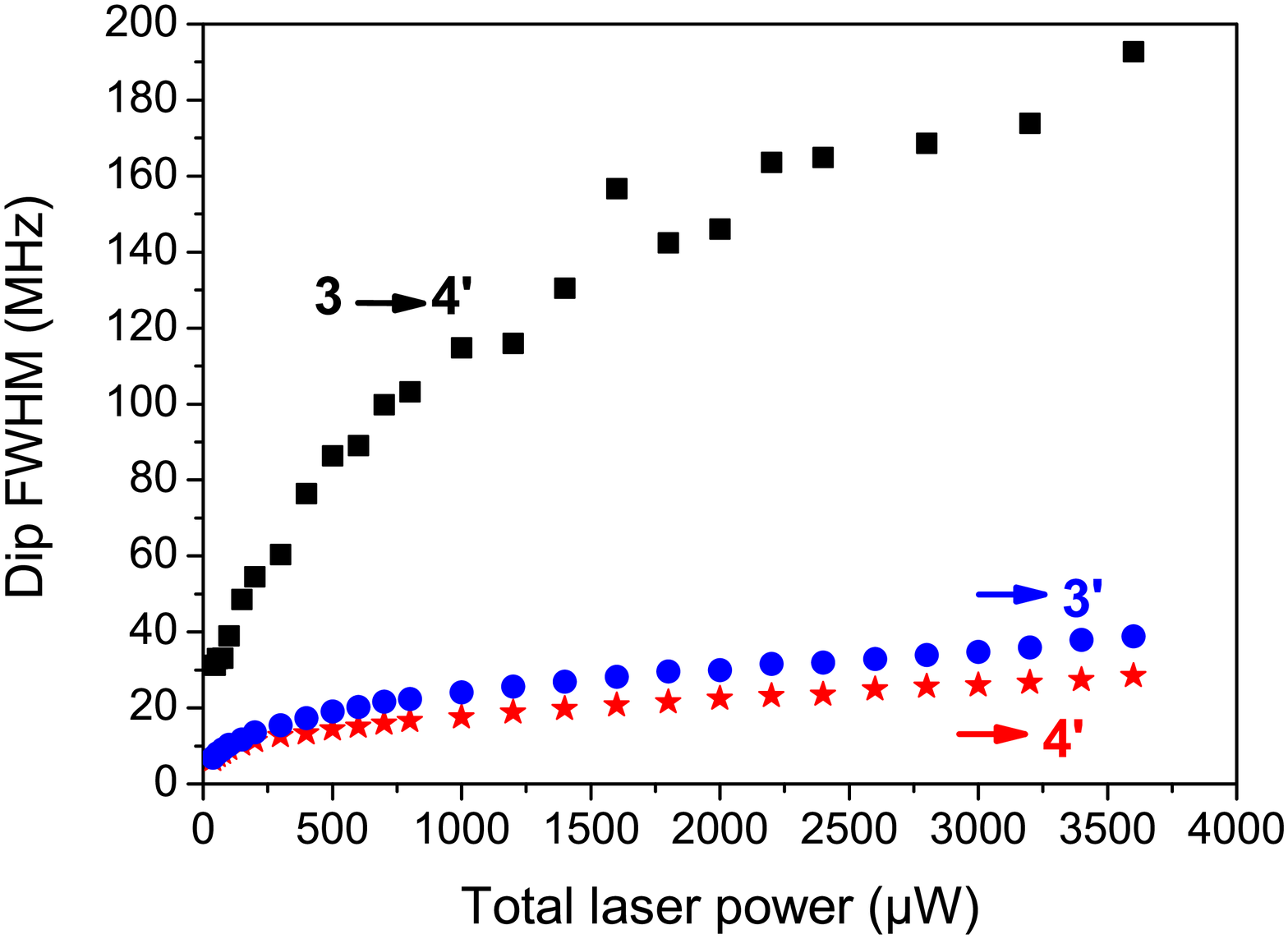}
\label{fig:fwhm-dip-P}} \hfill
\subfigure[]{\includegraphics[width=0.3\linewidth]{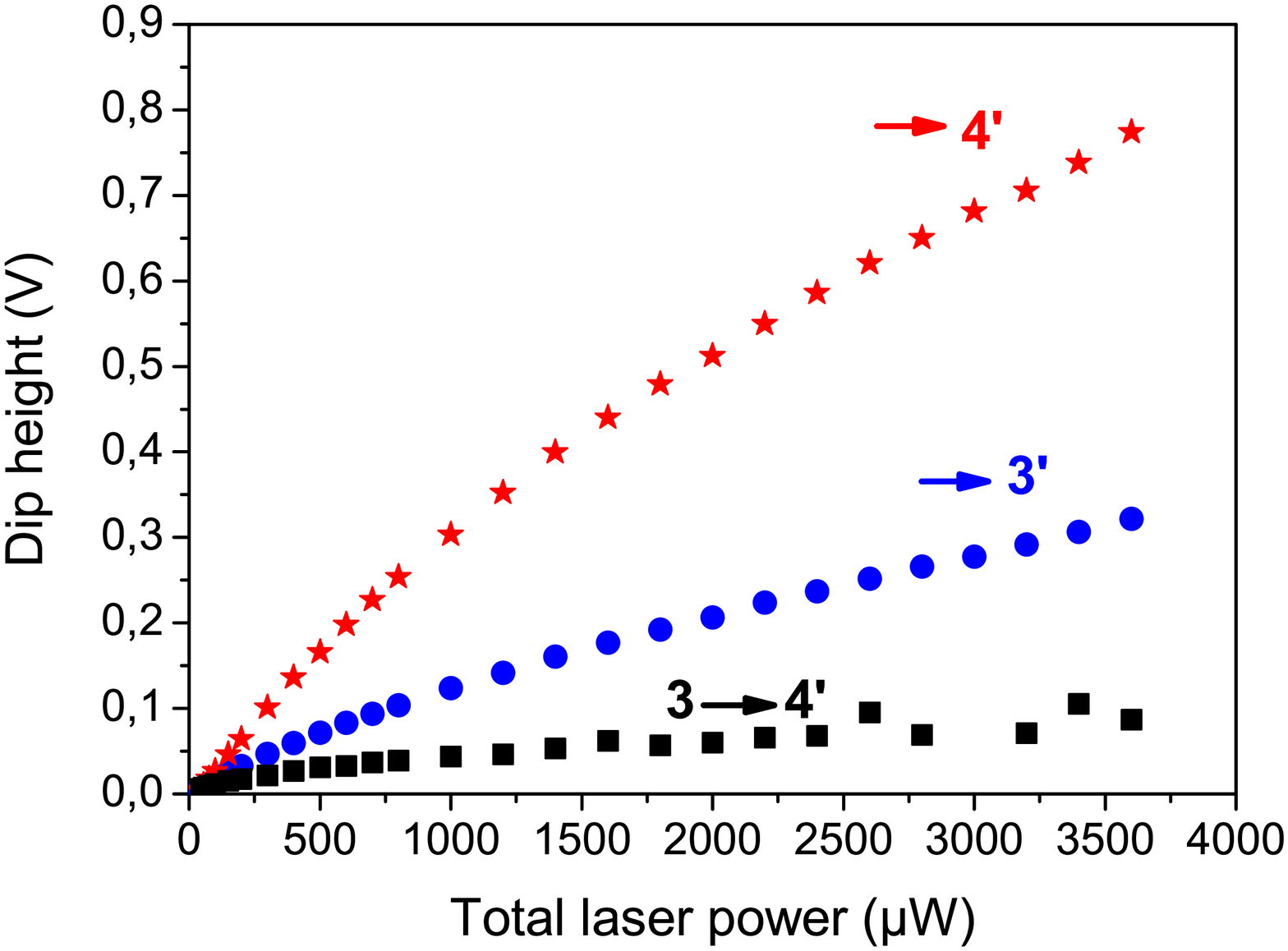}
\label{fig:signal-dip-P}} \hfill
\subfigure[]{\includegraphics[width=0.3\linewidth]{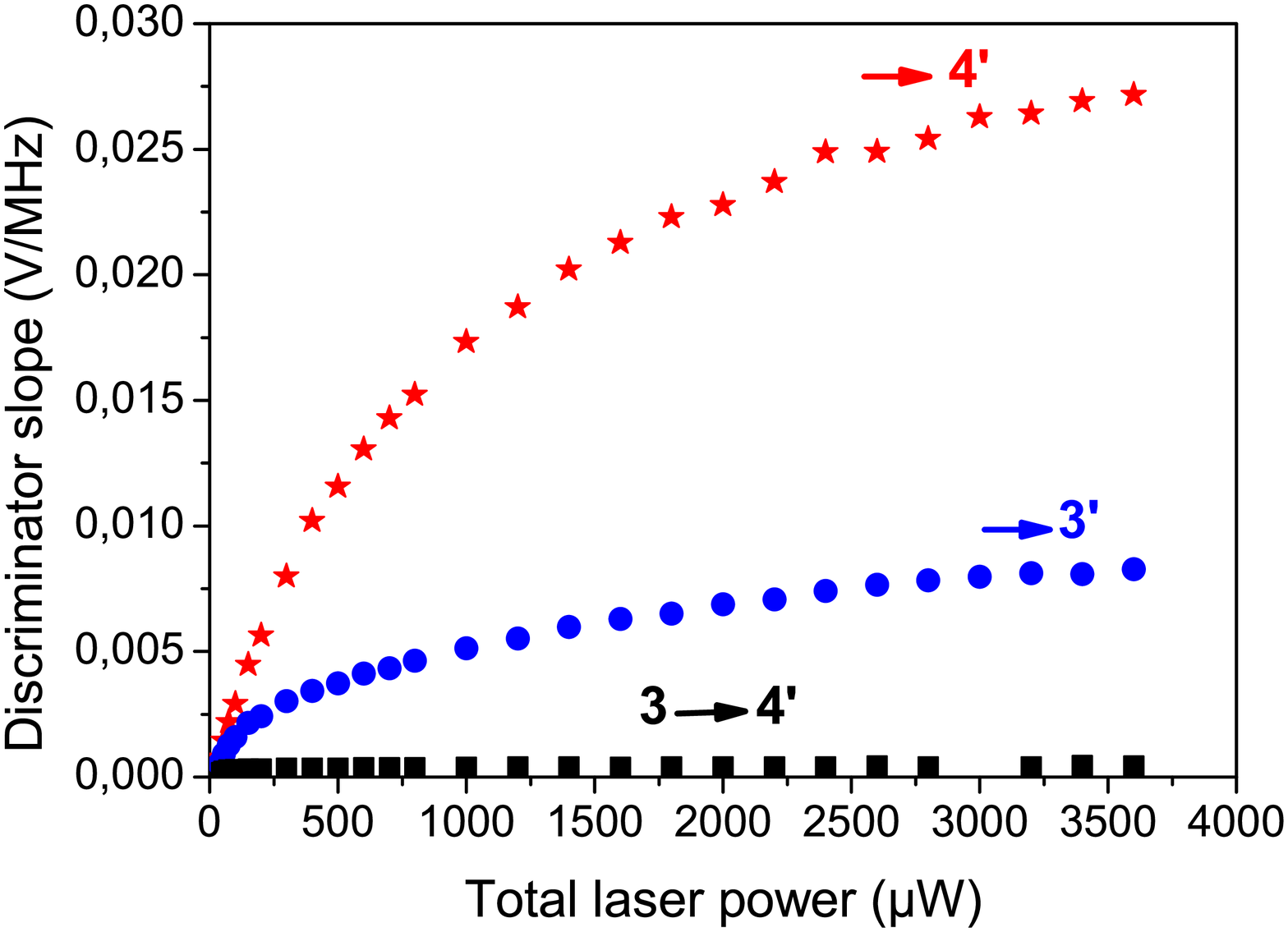}
\label{fig:slope-dip-P}}
 \caption{Full-width at half maximum (a), height (b) and height/FWHM ratio (c) of the Lorentzian dip signal detected in the bottom of the Doppler-broadened profile versus the total laser power incident in the cell. Squares: single-frequency regime (transition 3-4').  Dual-frequency regime (Stars: $\rightarrow F' =4$, Circles: $\rightarrow F' =3$).  }
\end{figure*}

Consider now atoms of a gas irradiated by this wave and the reflected counterpropagating wave. At non-zero values of the optical detuning $\Delta$, because of the Doppler shift, one travelling wave is resonant for atoms of the axial velocity class $v=\Delta/k$ ($k$ is the wave number) while the reflected wave is resonant for atoms of the velocity class $-v$. In that way, a fraction of atoms interacting with only one of the two waves is trapped in a dark state whatever the polarization of the other wave. At optical resonance the atoms of velocity class $v=0$ are resonant for both counter-propagating waves. When the polarization of the two fields is the same, dark states are common, less atoms can absorb the probe wave and the expected saturated absorption dip is observed. According to (\ref{eq:dsxy}), when the polarization vectors are orthogonal, atoms in the dark states of the pump beam become bright and absorbing for the probe, and vice-versa. In this case, the absorption of both beams is increased and the Lamb dip is reversed. A similar effect has already been investigated for linear and elliptical polarizations in the Hanle configuration \cite{Brazhnikov:2010, Brazhnikov:2011}. In this study, the monochromatic laser was tuned at resonance and the probe beam power was recorded versus the magnetic field. In Hanle configuration, the linewidth is limited by the CPT resonance width, inversely proportional to the Zeeman coherence lifetime. In the present experiment, the magnetic field is kept null. Then, the sub-Doppler resonance cancels when no more atoms interact with both waves simultaneously, i. e. when $\Delta>\Gamma$, with $\Gamma$ the linewidth of the optical transition. It can be shown that this CPT effect occurs when $F'=F-1$ \cite{Smirnov:1989}. It explains why we observe this phenomenon only on the ($|F = 4\rangle \rightarrow |F' = 3\rangle$) transition (see Fig. \ref{fig:spectres}).\\
The same effect occurs with a bichromatic beam carrying two optical frequencies frequency-split by the hyperfine-transition frequency. Since the Doppler-shift difference related to both frequencies is negligible in this case, hyperfine coherences between Zeeman sub-levels of both hyperfine states are also built. The previous argument is still valid. If the pump and probe beams have the same linear polarization, dark states are dark for both beams and the saturated absorption dip is observed. If pump and probe beams are orthogonally polarized, dark states are no more identical. For example, at $\Delta=0$, assuming an optically thin medium and equal field amplitudes, the low-frequency $\sigma^+$  component of pump beam ($\hat{x}$ polarized) and high frequency $\sigma^-$ component of probe beam ($\hat{y}$ polarized) induce the dark states:
\begin{eqnarray}
\begin{split}
\ket{DS_{x}} = d_{e,m;\,4,m-1}\ket{3,m+1}+d_{e,m;\,3,m+1}\ket{4,m-1},\\
\ket{DS_{y}} = d_{e,m;\,4,m-1}\ket{3,m+1}-d_{e,m;\,3,m+1}\ket{4,m-1},\\
\end{split}
\label{eq:dshf}
\end{eqnarray}
where $3 (4)$ holds for $F=3 \,(F=4)$, respectively, and $e$ for $F'=3$ or $F'=4$. Dark states of Eq. (\ref{eq:dshf}) are orthogonal. A higher absorption is expected at optical resonance. This occurs for Zeeman dark states and ($|\Delta F|=1,\, |\Delta m|=2$) dark states, but not for ($|\Delta F|=1,\, |\Delta m|=0$) dark states. In monochromatic and bichromatic cases, several state linear superpositions with  $|\Delta m|=2n, \, n\in  \mathbb{N}^*$, can coexist. However, since the atomic transit time across the beam is short, the steady-state is not reached and these dark states should be poorly populated.\\
In the case of crossed polarizations, we measured the evolution of the lorentzian dip versus the total laser power incident in the cell for both single-frequency and dual-frequency regimes. For each laser power, the absorption profiles of the Cs D$_1$ line spectrum are each fitted by the sum of a gaussian (Doppler broadened profile) and a lorentzian function (Doppler-free dip). Figures \ref{fig:fwhm-dip-P}, \ref{fig:signal-dip-P} and \ref{fig:slope-dip-P} report the full-width at half maximum (FWHM), height and height/FWHM ratio of the Lorentzian dip signal detected in the bottom of the Doppler-broadened profile versus the total laser power incident in the cell. We observe on Fig. \ref{fig:fwhm-dip-P} that the power broadening of the one-frequency dip is larger than in the two-frequency case. Simultaneously, the amplitude of the dip is dramatically increased in the dual-frequency case (see Fig. \ref{fig:signal-dip-P}). These observations are attributed to optical pumping effects. In the single-frequency case, the atomic system is an open system, causing the atoms to be pumped into the other hyperfine ground state and leading to a higher effective saturation parameter \cite{Lindvall:2007}. At the opposite, in the dual-frequency regime, the atomic system is closed, avoiding the loss of atoms. In the latter regime, a significant result is the important increase of the dip height with laser power, particularly for transitions connecting the ($F' = 4$) excited level. On Fig. \ref{fig:slope-dip-P}, we observe that the ratio signal/FWHM of the dip (slope of the frequency discriminator for laser frequency stabilization) is significantly increased in the dual-frequency regime. For a laser power of 500 $\mu$W, in the case of the transition to $F' = 4$, the dip-based frequency discriminator slope is about an order of magnitude higher than in the single-frequency case.\\
Laser frequency stabilization experiments were performed to compare single and dual-frequency regimes. For this purpose, two laser systems, named LS1 and LS2, were implemented. Both systems are mounted on distinct optical tables separated by about 2 meters and linked by an optical fiber. LS1 is similar to the system described in Fig. \ref{fig:setup}. LS2 uses a Distributed Bragg Resonator (DBR) laser and a phase EOM (PEOM) driven at 9.192 GHz. In this case, hyperfine CPT is induced by the carrier and one of the first-order sideband. The second first-order sideband is 9.192 GHz out of resonance and neglected here. We note that the gain on the frequency discriminator slope in the dual-frequency regime compared to the standard single frequency regime was found to be about the same for MZ EOM and PEOM-based setups. Different tests were performed. In test 1, each laser is in the free-running regime. In test 2, LS1 operates in the dual-frequency stabilization regime (DFSR) whereas LS2 is in the single-frequency stabilization regime (SFSR). For LS1, the MZ EOM is driven at 4.6 GHz with carrier suppression. The laser direct output (carrier) is shifted 4.6 GHz away from Cs resonance and the dual-frequency MZ EOM output beam is sent into the cell for laser stabilization. For LS2, the PEOM microwave modulation is turned off and the PEOM output beam is sent into the cell for laser stabilization onto a Cs transition. In test 3, both laser systems operate in the DFSR. LS1 is configured as in test 2 and the PEOM of LS2 is driven at 9.192 GHz. In all tests, the single-frequency direct output of each laser (before EOMs) is sent into a beam splitter cube. A laser beat-note at 4.596 GHz is detected by a fast photodiode, amplified and band-pass filtered. This signal is compared to a 4.621 GHz signal from a high-performance microwave synthesizer driven by a hydrogen maser. A final beat-note at 25 MHz is low-pass filtered and counted by a frequency counter (HP 53132A). Laser frequency is stabilized by modulating the laser injection current and demodulating it synchronously with a lock-in amplifier at the output of the photodiode. The bandwidth of laser frequency servos is about 1 kHz. Figure 4 reports the laser beat-note stability for each test.
\begin{figure}[h!]
\centering
\includegraphics[width=0.95\linewidth]{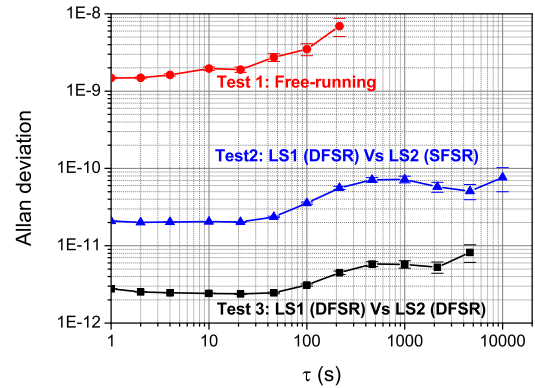}
\caption{Allan deviation of the laser beat-note frequency. Test1: free-running lasers, Test2: LS1 in the DFSR versus LS2 in the SFSR, Test 3: LS1 and LS2 are in the DFSR. For the latter, the total laser power incident in the cell is 500 $\mu$W.}
\label{fig:allan}
\end{figure}

In the free-running regime, the laser beat-note fractional frequency stability is $1.5 \times 10^{-9}$ at 1 s averaging time. In test 2 where one of the laser is stabilized through a standard single-frequency saturated absorption scheme, results are similar to those reported in \cite{Liu:IM:2012} and improved at the level of $2 \times 10^{-11}$ at 1 s and $3.5 \times 10^{-11}$ at 100 s. In test 3, where both lasers are stabilized through a dual-frequency Doppler-free spectroscopy setup, the laser beat-note frequency stability is improved by almost 10 yielding $3 \times 10^{-12}$ at 1 s and $5 \times 10^{-12}$ at 1000 s. For a laser power of 500 $\mu$W, this improvement is in correct agreement with the increased frequency discriminator slope reported in Fig. \ref{fig:slope-dip-P}. The bump at 500 s on Allan deviation plots is suspected to come from the room air conditioning cycle period. Degradation of the frequency stability at higher time scales could be explained by temperature variations, fiber perturbations or laser amplitude variations along time.\\

In conclusion, we reported the detection of natural-linewidth and high-signal inverted resonance dips through Doppler-free spectroscopy in a Cs vapor cell using a dual-frequency laser system. This technique has demonstrated excellent laser frequency stabilization at the level of 10$^{-12}$ at 1 s and 100 s integration time for a single laser, a gain of an order of magnitude compared to performances obtained with a conventional single-frequency laser system in similar conditions. This system can be used in various applications including CPT atomic clocks, laser spectroscopy, non-linear optics and laser-cooling experiments.

\section*{Acknowledgments}
This work has been funded by LabeX FIRST-TF, R\'egion de Franche-Comt\'e and the EMRP program (IND55 Mclocks). The EMRP is jointly funded by the EMRP participating countries within EURAMET and the European Union.

%Manual citation list


\begin{thebibliography}{1}

\bibitem{Schawlow}
A. L. Schawlow, Rev. Mod. Phys. \bf{54}\rm, 697-707 (1982).

\bibitem{Letokhov:1976}
V. S. Letokhov, "Saturation Spectroscopy", Chapter 4 of High Resolution Laser Spectroscopy (Topics in Applied Physics, Vol. 13, ed. K. Shimoda), Springer-Verlag, (1976).

\bibitem{VanierAudoin}
J. Vanier and C. Audoin, The quantum physics of atomic frequency standards, Adam-Hilger, Bristol (1989).

\bibitem{Romalis:Magneto}
D. Budker and M. Romalis, Nature Physics \bf{3}\rm, 227-234 (2007).

\bibitem{Chu:RMP:1998}
S. Chu, Rev. Mod. Phys. \bf{70}\rm, 3, 685 (1998).

\bibitem{Rovera:RSI:1994}
G. D. Rovera, G. Santarelli and A. Clairon, Rev. Sci. Instrum. \bf{65}\rm, 5, 1502-1505 (1994).

%\bibitem{Affolderbach:OLE:2005}
%C. Affolderbach and G. Mileti, Optics and Lasers in Engineering \bf{43}\rm, 291-302 (2005).

\bibitem{Affolderbach:RSI:2005}
C. Affolderbach and G. Mileti, Rev. Sci. Instr. \bf{76}\rm, 073108 (2005).

\bibitem{Baillard:OC:2006}
X. Baillard, A. Gauguet, S. Bize, P. Lemonde, P. Laurent, A. Clairon and P. Rosenbusch, Opt. Comm. \bf{266}\rm, 609-613 (2006).

\bibitem{Liu:IM:2012}
X. Liu and R. Boudot, IEEE Trans. Instr. Meas. \bf{61}\rm, 10, 2852-2855 (2012).

\bibitem{Liang:AO:2015}
W. Liang, V. S. Ilchenko, D. Eliyahu, E. Dale, A. A. Savchenkov, D. Seidel, A. B. Matsko and L. Maleki, Appl. Opt. \bf{54}\rm, 11, 3353-3359 (2015).

%\bibitem{Demtroder:Book}
%W. Demtroder, "Laser Spectrocopy: Basic concepts and instrumentation
%- Third edition," {\em Springer-Verlag ISBN 1439-2674}, Section
%5.4.5 (2002).

\bibitem{Pappas:1980}
P. G. Pappas, M. M. Burns, D. D. Hinshelwood, M. S. Feld, D. E. Murnick, Phys. Rev. A \bf{21}, \rm 1955 (1980).

\bibitem{Wynands:APB:1994}
O. Schmidt, K.-M. Knaak, R. Wynands and D. Meschede, Appl. Phys. B \bf{59}, \rm 167 (1994).

%\bibitem{Ishikawa}
%J. Ishikawa, F. Riehle, J. Helmcke, and C. Borde, Phys. Rev. A \bf{49}, \rm 4794 (1994).

\bibitem{McFerran:Arxiv:2016}
J. McFerran, arXiv \bf{1603.00970}, \rm (2016).

\bibitem{MAH:JAP:2015}
M. Abdel Hafiz and R. Boudot, Journ. Appl. Phys. \bf{118}, \rm 124903 (2015).

\bibitem{Knappe:2007}
S. Knappe, Emerging topics: MEMS Atomic Clocks, Comp. Microsys. \bf{3}, \rm 517-612 (2007).

\bibitem{Liu:PRA:2013}
X. Liu, J. M. M\'erolla, S. Gu\'erandel, C. Gorecki, E. De Clercq and R. Boudot, Phys. Rev. A \bf{87}, \rm 013416 (2013).

\bibitem{Arimondo:1996}
E. Arimondo, Prog. Opt. \bf{35}, \rm 257 (1996).

\bibitem{Beverini:1985}
N. Beverini, K. Ernst, M. Inguscio, anf F. Strumia, Appl. Phys. B \bf{37}, \rm 17 (1985).

\bibitem{Clercq:1989}
E. de Clercq and P. Mangin, Proc. European Freq. Time Forum EFTF, \rm 281, Besan\c con (1989).

\bibitem{Brazhnikov:2010}
D. V. Brazhnikov, A. V. Taichenachev, A. M. Tumaikin, V. I. Yudin, I. I. Ryabtsev, and V. M. Entin, JETP Lett. \bf{91}, \rm 625 (2010).

\bibitem{Brazhnikov:2011}
D. V. Brazhnikov, A. V. Taichenachev, and V. I. Yudin, Eur.Phys. J. D. \bf{63}, \rm 315 (2011).

\bibitem{Smirnov:1989}
V. S. Smirnov, A.M. Tumaikin, V. I. Yudin, J. Exp. Theor. Phys. \bf{69}, \rm 913 (1989).

\bibitem{Lindvall:2007}
T. Lindvall, I. Tittonen, J. Mod. Optics  \bf{54}, \rm 2779 (2007).




\end{thebibliography}
\end{document}